\documentclass[aps]{revtex4}
\usepackage{amsfonts}
\usepackage{amsmath}
\usepackage{amssymb,epsf}

\begin{document}

\title{Neutron stars in Einstein-$\Lambda$ gravity: the cosmological
constant effects}
\author{G. H. Bordbar$^{1,2}$\footnote{%
email address: ghbordbar@shirazu.ac.ir}, S. H. Hendi$^{1,3}$\footnote{%
email address: hendi@shirazu.ac.ir (corresponding author)} and B. Eslam Panah%
$^{1}$\footnote{%
email address: behzad.eslampanah@gmail.com} } \affiliation{$^1$
Physics Department and Biruni Observatory, College of Sciences,
Shiraz
University, Shiraz 71454, Iran\\
$^1$ Center for Excellence in Astronomy \& Astrophysics of Iran
(CEAAI-RIAAM)- Maragha, IRAN, P.O. Box: 55134-441, Iran\\
$^3$ Research Institute for Astronomy and Astrophysics of Maragha (RIAAM),
P.O. Box 55134-441, Maragha, Iran}

\begin{abstract}
Regarding a $d-$dimensional spherically symmetric line element in the
context of Einstein-$\Lambda$ gravity, the hydrostatic equilibrium equation
of stars is obtained. Then, by using the lowest order constrained
variational (LOCV) method with the AV$18$ potential and employing
microscopic many body calculations in the modern equation of state, the
structure properties of neutron stars are investigated. Regardless of
cosmological point of view and considering arbitrary positive and negative
values of the cosmological constant, the maximum mass of the neutron stars
and their corresponding radius in $4$-dimensions are computed. The results
show that there is an upper limit for the maximum mass of neutron star for
positive cosmological constant ($M_{\max }\leq 1.68M_{\odot }$). On the
other hand, it is shown that the Einstein gravity cannot explain the
structure of neutron star with negative $\Lambda$. Other properties of
neutron stars such as; the Schwarzschild radius, average density,
compactness and Buchdahl- Bondi bound are studied. In addition, by using the
Buchdahl-Bondi bound for neutron stars, stability of these stars is
investigated. Finally, the dynamical stability is investigated and shown
that the neutron stars follow the dynamical stability in this gravity.
\end{abstract}

\maketitle

\section{Introduction}

Most of the solar system phenomena such as the Mercury precession can be
successfully explained in the context of Einstein gravity, but when we
intend to study beyond the solar system or strong gravity regimes, this
theory encounters some problems. Accelerated expansion of our universe is
one of the main challenging subjects of cosmologists \cite%
{Perlmutter1,Perlmutter2,Perlmutter3}, whereas Einstein gravity cannot
properly explain it. In order to interpret this expansion, some various
candidates have been proposed by many authors. Among various alternatives,
one may point out the modified gravities like Lovelock gravity \cite%
{Lovelock1,Lovelock2,Dehghani1,Dehghani2}, brane world cosmology \cite%
{Brax1,Brax2,Brax3}, scalar-tensor theories \cite%
{Brans1,Brans2,Brans3,Brans4,Brans5,Harada1,Harada2,Harada3,Harada4,Harada5}
and $F(R)$ gravity \cite%
{Souza1,Souza2,Souza3,Souza4,Souza5,Souza6,Cooney1,Cooney2,Cooney3,Cooney4,Cooney5}%
.

From the cosmological point of view, the dark energy is an unknown form of
energy which is hypothesized to permeate the spacetime, tending to
accelerate the expansion of the universe \cite{Peebles}. The dark energy is
the most accepted hypothesis to explain the observations since the 1990s
indicating that the universe is expanding at an accelerating rate. Another
model of these interesting candidates is related to the cosmological
constant. In this theory, the cosmological constant may add to the Einstein
Lagrangian \cite{Padmanabhan1,Padmanabhan2} for explaining such
acceleration. The cosmological constant is the simplest possible form of
dark energy, which leads to the current standard model of cosmology known as
the $\Lambda $-CDM model and provides a good fit to many cosmological
observations.

From the astrophysical point of view, stars reach the equilibrium state due
to the balance between gravitational force and internal pressure. Therefore,
in order to study the structure of stars, we should obtain a suitable
hydrostatic equilibrium equation (HEE) and then we can study the structure
properties of stars. The neutron and quark stars are in the category of
celestial objects with high mass, high density and small radius, and
therefore, they are called compact objects. Due to this fact, we need to
take into account the effects of general relativity (spacetime curvature)
for investigating their structures. The first HEE of stars in $4$%
-dimensional Einstein gravity has been studied by Tolman, Oppenheimer and
Volkoff (TOV) \cite{TOV1,TOV2,TOV3}. Also, the physical characteristics of
stars using TOV equation have been investigated by many authors \cite%
{Silbar1,Silbar2,Silbar3,Silbar4,Silbar5,Silbar6,Silbar7,Silbar8}. Recently,
survey of compact objects in different gravity has attracted the attention
of researchers, in particular, the models of modified gravity which involve
higher curvature terms or a scalar field \cite{Various gravity1,Various
gravity2,Various gravity3,Various gravity4,Various gravity5,Various
gravity6,Various gravity7,Various gravity8,Various gravity9,Various
gravity10,Hendi2015}. On the other hand, if one is interested to study the
structure and evolution of stars in different theories of gravity with
various sources, the new HEE must be obtained. Therefore, in recent years,
the generalizations and modifications of TOV equation have been investigated
in various theories of gravity, such as: dilaton gravity \cite{Hendi2015},
gravity's rainbow \cite{HendiBEP}, massive gravity \cite{HendiBEmassive}, $%
F(R)$ and $F(G)$\ gravities \cite%
{Astashenok2015aAstashenok2015b,Momeni1,Momeni2,Momeni3,Momeni4} and p
ost-Newtonian theory \cite{Glampedakis} (see also Refs. \cite%
{Boyadjiev,Vinayaraj1,Vinayaraj2,Vinayaraj3,Astashenok2013a,Astashenok2013b,Astashenok2013c,Astashenok2013d,Lemos1,Lemos2,Lemos3,Lemos4,Green1,Green2}
for more details).

The outline of our paper is as follows. In the next section, we consider a
spherical symmetric metric and obtain the HEE in the Einstein-$\Lambda $
gravity for arbitrary dimensions ($d\geq 3$). Then, we investigate stability
and energy conditions for the equation of state of neutron star matter in $4$%
-dimensions. Also, we study effects the cosmological constant on maximum
mass and its radius of the neutron stars. Next, we examine effects of the
cosmological constant on other properties of neutron stars such as;
Schwarzschild radius, average density, compactness, Buchdahl-Bondi bound and
dynamical stability. We finish our paper with some closing remarks.

\section{HEE in the presence of cosmological constant}

As first step, we regard $d-$dimensional Einstein$-\Lambda$ gravity to
obtain its related HEE. The action of the Einstein gravity with the
cosmological constant in arbitrary dimensions, $d$, is given by
\begin{equation}
I_{G}=-\frac{1}{16\pi }\int_{\mathcal{M}}d^{d}x\sqrt{-g}\left(R-2\Lambda
\right)+I_{Matt},  \label{actionEN}
\end{equation}%
where $R$ and $\Lambda$ are, respectively, the Ricci scalar and the
cosmological constant in $d-$dimensions and $I_{Matt}$ is the action of
matter field. Varying the action (\ref{actionEN}) with respect to the metric
tensor $g_{\mu }^{\nu }$, the equation of motion can be written as
\begin{equation}
R_{\mu }^{\nu }+\frac{1}{2}Rg_{\mu }^{\nu }+\Lambda g_{\mu }^{\nu
}=K_{d}T_{\mu }^{\nu },  \label{EqEN}
\end{equation}%
where $K_{d}=\frac{8\pi G_{d}}{c^{4}}$, $G_{d}$ is the gravitational
constant in $d$-dimensions and define as $G_{d}=GV_{d-4}$, where $G$ denotes
the four dimensional gravitational constant and $V_{d-4}$ is the volume of
extra dimensions. Also, the constant $c$ is the speed of light in the
vacuum. In addition, $R_{\mu }^{\nu }$ and $T_{\mu }^{\nu }$ are the
symmetric Ricci tensor and energy-momentum tensors, respectively.

Here, we want to obtain the static solutions of Eq. (\ref{EqEN}). For this
purpose, we assume a spherical symmetric spacetime in the following form
\begin{equation}
ds^{2}=f(r)dt^{2}-\frac{dr^{2}}{g(r)}-r^{2}d\Omega _{k}^{2},  \label{Metric}
\end{equation}%
where $f(r)$ and $g(r)$ are the unknown metric functions of radial
coordinates and $d\Omega _{k}^{2}$ is the line element of unit $(d-2)$%
-dimensional sphere
\begin{equation}
d\Omega _{k}^{2}=d\theta
_{1}^{2}+\sum\limits_{i=2}^{d-2}\prod\limits_{j=1}^{i-1}\sin ^{2}\theta
_{j}d\theta _{i}^{2}.
\end{equation}

On the other hand, the energy-momentum tensor for a perfect fluid is
\begin{equation}
T^{\mu \nu }=\left( c^{2}\epsilon +P\right) U^{\mu }U^{\nu }-Pg^{\mu \nu },
\label{EMTensorEN}
\end{equation}%
where $\epsilon $ and $P$ are density and pressure of the fluid which are
measured by local observer, respectively, and $U^{\mu }$ is the fluid
four-velocity. Using Eqs. (\ref{EqEN}) and (\ref{EMTensorEN}) and the metric
introduced in Eq. (\ref{Metric}), we can obtain the components of
energy-momentum for $d$-dimensions as follows
\begin{equation}
T_{0}^{0}=\epsilon c^{2}~\ \ \ \ \&~\ \ \ \
T_{1}^{1}=T_{2}^{2}=T_{3}^{3}=...=T_{d-1}^{d-1}=-P.  \label{4dim}
\end{equation}

We consider the metric (\ref{Metric}) and Eq. (\ref{4dim}) for the perfect
fluid and obtain the components of Eq. (\ref{EqEN}) in the following forms
\begin{eqnarray}
&&K_{d}c^{2}r^{2}\epsilon =\Lambda r^{2}+\frac{d_{2}d_{3}}{2}\left(
1-g\right) -\frac{d_{2}}{2}rg{^{\prime }},  \label{1} \\
&&  \notag \\
&&K_{d}fr^{2}P=-\Lambda r^{2}f-\frac{d_{2}d_{3}}{2}\left( 1-g\right) f+\frac{%
d_{2}}{2}rg{f{^{\prime }},}  \label{2} \\
&&  \notag \\
&&4K_{d}f^{2}rP=-4\Lambda rf^{2}-\frac{2d_{3}d_{4}\left( 1-g\right) f^{2}}{r}%
+2d_{3}\left( gf\right) {^{\prime }}f-rgf{^{\prime 2}}+r\left[ g{^{\prime }}f%
{^{\prime }+2g}f{^{\prime \prime }}\right] f,  \label{33}
\end{eqnarray}%
where $d_{i}=d-i$ and also, $f$, $g$, $\epsilon $ and $P$ are functions of $%
r $. We note that the prime and double prime are, respectively, the first
and second derivatives with respect to $r$.

Using Eqs. (\ref{1}) - (\ref{33}) and after some calculations, we obtain
\begin{equation}
\frac{dP}{dr}+\frac{\left( c^{2}\epsilon +P\right) f{^{\prime }}}{2f}=0.
\label{extraEQ}
\end{equation}

Now, we obtain $f{^{\prime }}$ from Eq. (\ref{2}) as follows
\begin{equation}
f{^{\prime }}=\frac{2\left[ r^{2}\left( \Lambda +K_{d}P\right) +\frac{%
d_{2}d_{3}}{2}\left( 1-g\right) \right] f}{rgd_{2}}.  \label{diff(r)}
\end{equation}

Then, we calculate $g(r)$ by using Eq. (\ref{1}) with the following form
\begin{equation}
g(r)=1+\frac{2\Lambda }{d_{1}d_{2}}r^{2}-\frac{c^{2}K_{d}M\Gamma \left(
\frac{d_{1}}{2}\right) }{d_{2}\pi ^{d_{2}/2}r^{d-3}},  \label{4g(r)}
\end{equation}%
where $M=\int \frac{2\pi ^{d_{1}/2}}{\Gamma (d_{1}/2)}r^{d_{2}}\epsilon
(r)dr $ and $\Gamma $ is the gamma function, in which $\Gamma (1/2)=\sqrt{%
\pi }$, $\Gamma (1)=1$ and $\Gamma (x+1)=x\Gamma (x)$.

By considering Eqs. (\ref{diff(r)}) and (\ref{4g(r)}), and inserting them in
Eq. (\ref{extraEQ}), we can extract the HEE of Einstein-$\Lambda $ gravity
in $d$-dimensions as
\begin{equation}
\frac{dP}{dr}=\frac{\left[ \frac{d_{1}d_{3}\Gamma (\frac{d_{1}}{2})}{4\pi
^{d_{1}/2}}c^{2}K_{d}M+r^{d_{1}}\left( \Lambda +\frac{d_{1}}{2}K_{d}P\right) %
\right] }{r\left[ -\Lambda r^{d_{1}}+d_{1}\left( \frac{\Gamma (\frac{d_{1}}{2%
})}{2\pi ^{^{d_{1}/2}}}c^{2}K_{d}M-\frac{d_{2}}{2}r^{d_{3}}\right) \right] }%
\left( c^{2}\epsilon +P\right) ,  \label{TOV}
\end{equation}%
where for $4$-dimensional case ($d=4$), Eq. (\ref{TOV}) reduces to the
following equation
\begin{equation}
\frac{dP}{dr}=\frac{\left[ 3c^{2}GM+r^{3}\left( \Lambda c^{4}+12\pi
GP\right) \right] }{c^{2}r\left[ 6GM-c^{2}r\left( \Lambda r^{2}+3\right) %
\right] }\left( c^{2}\epsilon +P\right) .  \label{4TOV}
\end{equation}

Also, Eq. (\ref{4TOV}) leads to primitive TOV equation for vanishing $%
\Lambda $ \cite{TOV1,TOV2,TOV3}. It is evident that for $d=3$, Eq. (\ref{TOV}%
) reduces to the HEE obtained by Diaz \cite{Diaz}. As a result, we have
obtained a global form of HEE for arbitrary dimensions ($d\geq 3$) in Eq. (%
\ref{TOV}).

\section{Structure properties of neutron star}

\subsection{Equation of state of neutron star matter}

Obtaining equation of state for neutron stars enables one to study its
properties. The interior region of the neutron star is a mixture of
neutrons, protons, electrons and muons. These constituents are in charge
neutrality and beta equilibrium conditions (beta-stable matter) \cite%
{Shapiro}. Recently, the microscopic constrained variational calculations
based on the cluster expansion of the energy functional has been employed to
obtain the equation of state of neutron star matter \cite%
{Bordbar1,Bordbar2,Hendi2015}. The two-nucleon potentials which are used in
these studies are the modern Argonne AV$18$ \cite{Wiringa} and charged
dependent Reid-$93$ \cite{Stoks}. One of the advantages of these methods is
that there is no need for any free parameter in formalism, and they result
into a good convergence. Also, the results in this method are more accurate
comparing to other semi-empirical parabolic approximation methods. This is
due to a microscopic computation of asymmetry energy which is carried on for
the asymmetric nuclear matter calculations. In fact, it is necessary to have
a microscopic calculation with the modern nucleon-nucleon potentials which
is isospin projection ($T_{z}$) dependant \cite{Bordbar1998}. Here, we are
interested in using the lowest order constrained variational (LOCV) method
with the AV$18$ potential \cite{Bordbar1,Bordbar2} for calculating the
modern equation of state for neutron star matter and investigating some
physical properties of neutron star structure. We use the LOCV method to
calculate the energy of our system. The LOCV method is a useful tool for
determination of the properties of neutron, nuclear and asymmetric nuclear
matter at finite and zero temperature \cite%
{Modarres,ModarresI,Howes,Owen,ModarresII}. It is notable that, the LOCV
method is a fully self-consistent formalism and it does not bring any free
parameters into calculation. It employs a normalization constraint to keep
the higher order term as small as possible \cite{Owen,BordbarM1997}. The
functional minimization procedure represents an enormous computational
simplification over unconstrained methods that attempt to go beyond lowest
order.

A suitable trial many-body wave function is
\begin{equation}
\psi =F\phi ,
\end{equation}%
where $\phi $ is the uncorrelated ground-state wave function of $N$
independent neutrons, and $F$ is a proper $N$-body correlation function.
Applying Jastrow approximation \cite{Jastrow}, $F$ is rewritten by
\begin{equation}
F=S\prod_{i>j}f(ij),
\end{equation}%
where $S$ is a symmetrizing operator. Also, we consider a cluster expansion
of the energy functional up to the two-body term%
\begin{equation}
E([f])=\frac{1}{N}\frac{\langle \psi |H|\psi \rangle }{\langle \psi |\psi
\rangle }=E_{1}+E_{2},
\end{equation}

\begin{figure}[tbp]
$%
\begin{array}{c}
\epsfxsize=7cm \epsffile{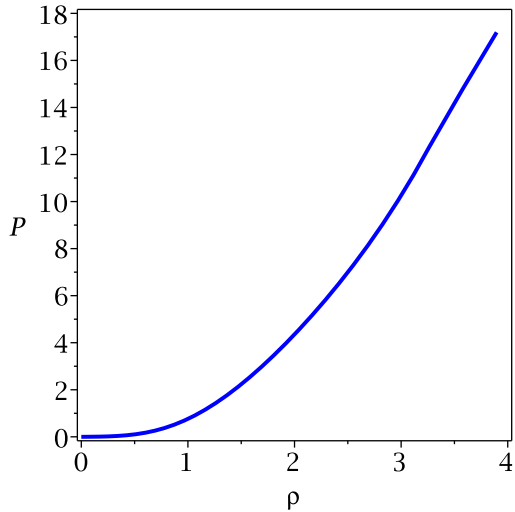}%
\end{array}
$%
\caption{Equation of state of neutron star matter (pressure, $P$ ($10^{35}$
erg/$cm^{3}$) versus density, $\protect\epsilon $ ($10^{15}$ gr/$cm^{3}$)).}
\label{Fig1}
\end{figure}

The energy per particle up to the two-body term is
\begin{equation}
E([f])=E_{1}+E_{2},
\end{equation}%
where $E_{1}=\sum_{i=+,-}\frac{3}{5}\frac{\hbar ^{2}k_{F}^{(i)^{2}}}{2m}%
\frac{\epsilon ^{(i)}}{\epsilon }$ and $E_{2}=\frac{1}{2N}\sum_{ij}\langle
ij|\nu (12)|ij-ji\rangle $ are one-body and two-body energy term,
respectively. The operator $\nu (12)$ has been given in Ref. \cite%
{Bordbar044310} (see Refs. \cite{Bordbar1998,BordbarB1,BordbarB2} for more
details). Our obtained equation of state of neutron star matter is shown in
Fig. \ref{Fig1}.

In order to obtain more details for equation of state of neutron star matter
introduced in this paper, we study both stability and energy conditions in
the following subsections.

\subsubsection{Stability}

In order to investigate the stability of the introduced equation of state
for a physically acceptable model, one expects that, the speed of sound ($v$%
) be less than the speed of light ($c$) \cite{Herrera1,Herrera2}. This
stability condition can be written in the following explicit form
\begin{equation}
0\leq v^{2}=\left( \frac{dP}{d\epsilon }\right) \leq c^2.
\end{equation}

By using the above condition and Fiq. \ref{Fig1}, and comparing them with
diagrams related to sound speed-radius relationship plotted in Ref. \cite%
{Bordbar2006} (see Fig. 2 in Ref. \cite{Bordbar2006}, for more details), it
is evident that this equation of state satisfies the inequality $0\leq
v^{2}\leq c^2$. Therefore, the introduced equation of state is stable.

\subsubsection{Energy Conditions}

One of interesting properties of the mentioned equations of state is related
to its central density. It is notable that the obtained central density of
this equations of state is about $3.9\times 10^{15}\ g~cm^{-3}$ and this
density is larger than the normal nuclear density $\epsilon _{0}=2.7\times
10^{14}g~cm^{-3}$ \cite{WiringaFF}.

Now, we want to study the energy conditions. In this regard, we investigate
null energy condition (NEC), weak energy condition (WEC), strong energy
condition (SEC) and dominant energy condition (DEC) at the center of the
neutron stars. These energy conditions can be written in the following
explicit forms
\begin{eqnarray}
NEC &\rightarrow &\ P_{c}+\epsilon _{c}\geq 0,  \label{11} \\
WEC &\rightarrow &\ P_{c}+\epsilon _{c}\geq 0,\ \ \ \&\ \ \epsilon _{c}\geq
0,  \label{22} \\
SEC &\rightarrow &\ P_{c}+\epsilon _{c}\geq 0,\ \ \ \&\ \ 3P_{c}+\epsilon
_{c}\geq 0, \\
DEC &\rightarrow &\ \epsilon _{c}>\left\vert P_{c}\right\vert ,  \label{44}
\end{eqnarray}
where $P_{c}$ and $\epsilon_{c}$ are the pressure and density at the center
of neutron star ($r=0$), respectively. Considering Fig. \ref{Fig1} and the
above conditions (\ref{11})-(\ref{44}), our results are presented in Table %
\ref{tab11}.

\begin{table*}[tbp]
\caption{Energy conditions for neutron stars.}
\label{tab11}
\begin{center}
\begin{tabular}{cccccc}
\hline\hline
$\epsilon _{0}\ (10^{9}kg/cm^3)$ & $P_{0}\ (10^{9}kg/cm^3)$ & $NEC$ & $WEC$
& $SEC$ & $DEC$ \\ \hline\hline
$3895.99$ & $1910.39$ & $\checkmark$ & $\checkmark$ & $\checkmark$ & $%
\checkmark$ \\ \hline\hline
&  &  &  &  &
\end{tabular}%
\end{center}
\end{table*}

According to the Table \ref{tab11}, we observe that all the energy
conditions are satisfied. Therefore, the equation of state of neutron star
matter introduced in this paper are suitable and consistent with all
versions of energy conditions.

The equation of state of neutron star matter satisfied both stability and
energy conditions. Now, we focus on investigation of maximum mass and radius
for the neutron stars in Einstein-$\Lambda $ gravity.

\subsection{Properties of neutron stars in Einstein-$\Lambda $ gravity}

Considering the maximum gravitational mass of a neutron star for stability
against collapse into a black hole, one is able to recognize differences
between neutron stars and black holes. In other words, there is a critical
maximum mass in which for smaller than such mass, degenerate pressure
originated from the nucleons prevents an object from becoming a black hole
\cite{Shapiro}. Therefore, obtaining the maximum gravitational mass of
neutron stars is of great importance in astrophysics. Due to many errors in
direct ways of measuring the neutron star mass by observations of the X-ray
pulsars and X-ray bursts, one is not able to obtain an accurate mass for
these stars. On the other hand, using the binary radio pulsars \cite%
{Weisberg1,Weisberg2,Weisberg3,Weisberg4}, one may obtain highly accurate
results for the mass of neutron stars.

Now, by employing the equation of state of neutron star matter presented in
Fig. \ref{Fig1} and numerical approach for integrating the HEE obtained in
Eq. (\ref{TOV}), we obtain the maximum mass of neutron stars. The neutron
star mass and radius depend on the central mass density ($\epsilon _{c}$).
We can consider the boundary conditions $P(r=0)=P_{c}$ and $m(r=0)=0$, and
integrate Eq. (\ref{TOV}) outward to a radius $r=R$ in which $P$ vanishes
for selecting an $\epsilon _{c}$. This leads to the neutron star radius $R$
and mass $M=m(R)$.

In Ref. \cite{Bordbar2006}, the Einstein gravity without cosmological
constant has been investigated, and maximum mass of neutron stars has been
obtained using the modern equations of state of neutron star matter derived
from microscopic calculations. It was shown that the maximum mass of neutron
star is about $1.68M_{\odot }$. In the present paper, we are going to
consider the cosmological constant and obtain the maximum mass for neutron
stars by employing the modern equations of state and study other properties
of these stars. Our results are presented in Tables \ref{tab2} and \ref{tab3}%
.

The value of the cosmological constant and its fine tuning is open questions
for scientists. In addition, its consistent values in various scenarios are
different and therefore, it is allowed to regard it as a free parameter. On
the other hand, from the standpoint of cosmology its order of magnitude may
be approximately $10^{-52}$ $m^{-2}$. Such order of magnitude may be valid
in a special cosmological model with large scale structure. Meanwhile, in
order to regard the effects of $\Lambda$ in local scales (such as near the
black holes, neutron stars and other massive objects), it is not necessary
to follow the fine tuning values of $\Lambda$ is large scales. Here, we
regard the value $10^{-52}$ $m^{-2}$ and also other values (at least as a
toy model) to obtain the effects of cosmological constant on the neutron
star structure. According to our results, the cosmological constant has no
significant effect when we consider its value identical to $10^{-52}$ $%
m^{-2} $ (see Table \ref{tab2}, for more details). Therefore, in order to
examine its effects, we should consider a (an ad hoc) model in which its
value is less than $10^{-14}m^{-2}$.
\begin{table*}[tbp]
\caption{Einstein-$\Lambda $ gravity for $4$-dimensions.}
\label{tab2}
\begin{center}
\begin{tabular}{cccccccc}
\hline\hline
$\Lambda(m^{-2})$ & $\frac{{M_{max}}}{(M_{\odot})}$ & $R(km)$ & $R_{Sch}(km)$
& $\overline{\epsilon }$ $(10^{15}g$ $cm^{-3})$ & $\eta (10^{-1})$ & $z$ & $%
M_{BB}\ (M_{\odot})$ \\ \hline\hline
$0$ & $1.68$ & $8.42$ & $4.95$ & $1.34$ & $5.88$ & $0.56$ & $2.54$ \\ \hline
$1.00\times 10^{-50}$ & $1.68$ & $8.42$ & $4.95$ & $1.34$ & $5.88$ & $0.56$
& $2.54$ \\ \hline
$1.00\times 10^{-52}$ & $1.68$ & $8.42$ & $4.95$ & $1.34$ & $5.88$ & $0.56$
& $2.54$ \\ \hline
$1.00\times 10^{-54}$ & $1.68$ & $8.42$ & $4.95$ & $1.34$ & $5.88$ & $0.56$
& $2.54$ \\ \hline\hline
&  &  &  &  &  &  &
\end{tabular}%
\end{center}
\end{table*}
\begin{table*}[tbp]
\caption{Einstein-$\Lambda $ gravity for $4$-dimensions.}
\label{tab3}
\begin{center}
\begin{tabular}{cccccccc}
\hline\hline
$\Lambda(m^{-2})$ & $\frac{{M_{max}}}{(M_{\odot})}$ & $R(km)$ & $R_{Sch}(km)$
& $\overline{\epsilon }$ $(10^{15}g$ $cm^{-3})$ & $\eta (10^{-1})$ & $z$ & $%
M_{BB}\ (M_{\odot})$ \\ \hline\hline
$1.00\times 10^{-11}$ & $0.78$ & $6.65$ & $2.30$ & $1.26$ & $3.46$ & $0.24$
& $2.00$ \\ \hline
$5.00\times 10^{-12}$ & $1.12$ & $7.47$ & $3.30$ & $1.27$ & $4.42$ & $0.34$
& $2.25$ \\ \hline
$1.00\times 10^{-12}$ & $1.56$ & $8.25$ & $4.60$ & $1.32$ & $5.57$ & $0.50$
& $2.49$ \\ \hline
$5.00\times 10^{-13}$ & $1.62$ & $8.34$ & $4.77$ & $1.33$ & $5.72$ & $0.53$
& $2.51$ \\ \hline
$1.00\times 10^{-13}$ & $1.67$ & $8.40$ & $4.92$ & $1.34$ & $5.86$ & $0.55$
& $2.53$ \\ \hline
$1.00\times 10^{-14}$ & $1.68$ & $8.42$ & $4.95$ & $1.34$ & $5.88$ & $0.56$
& $2.54$ \\ \hline
$1.00\times 10^{-15}$ & $1.68$ & $8.42$ & $4.95$ & $1.34$ & $5.88$ & $0.56$
& $2.54$ \\ \hline\hline
&  &  &  &  &  &  &
\end{tabular}%
\end{center}
\end{table*}

Based on Table \ref{tab3}, considering the cosmological constant as a
positive quantity, the maximum mass of neutron star decreases as $\Lambda $
increases. In other words, the maximum mass of neutron stars with the
positive values of cosmological constant are in the range $M_{\max }\leq
1.68M_{\odot }$. Also, when we consider the cosmological constant to be very
small ($\Lambda \leq 10^{-14}$), the maximum mass reduces to results
obtained in the Einstein gravity \cite{Bordbar2006}. In other words, for $%
\Lambda \leq 10^{-14}$ this constant does not affect the maximum mass and
other properties. So, considering the obtained value for the cosmological
constant from the cosmological perspective ($\Lambda \simeq 10^{-52}m^{-2}$%
), this quantity does not affect the structure of neutron stars.

\begin{figure}[tbp]
$%
\begin{array}{c}
\epsfxsize=7cm \epsffile{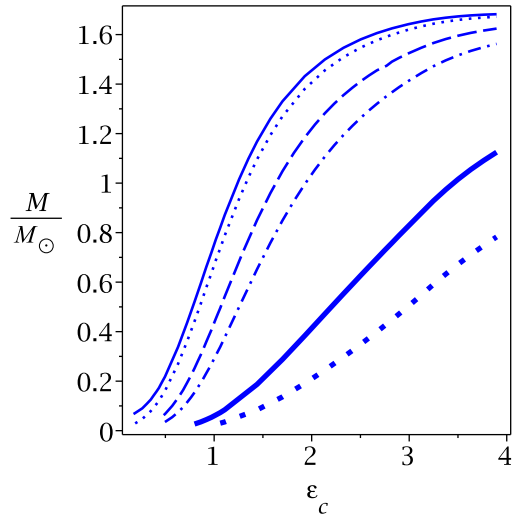}%
\end{array}
$%
\caption{Maximum mass of neutron star versus central mass density $\protect%
\epsilon _{c}$ ($10^{15}$gr/$cm^{3}$), for $\Lambda =1\times 10^{-14}$
(continuous line), $\Lambda =1\times 10^{-13}$ (dotted line), $\Lambda
=5\times 10^{-13}$ (dashed line) , $\Lambda =1\times 10^{-12}$
(dashed-dotted line), $\Lambda =5\times 10^{-12}$ (bold line) and $\Lambda
=1\times 10^{-11}$ (bold-dotted line).}
\label{Fig2}
\end{figure}

\begin{figure}[tbp]
$%
\begin{array}{cc}
\epsfxsize=7cm \epsffile{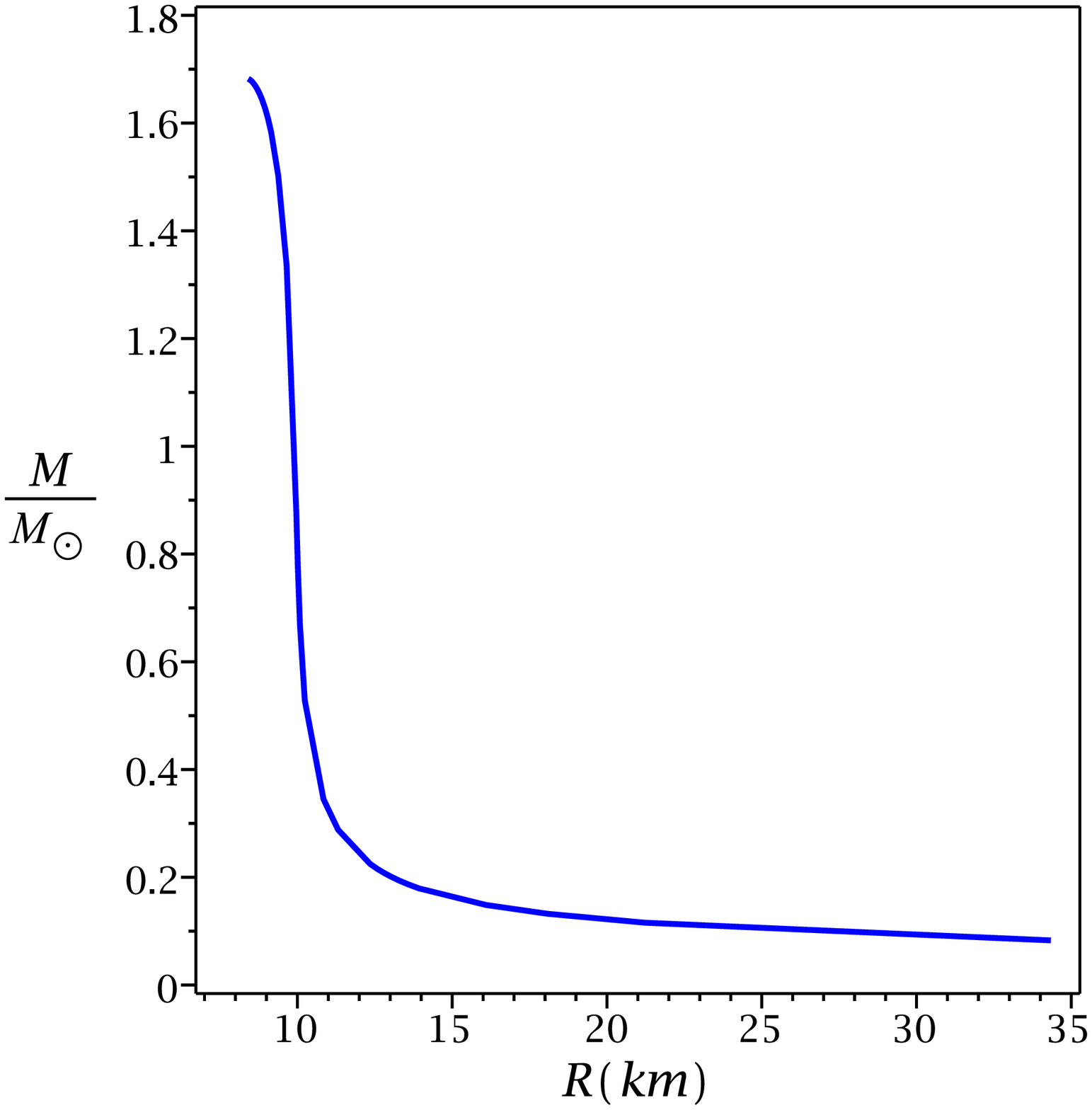} & \epsfxsize=7cm %
\epsffile{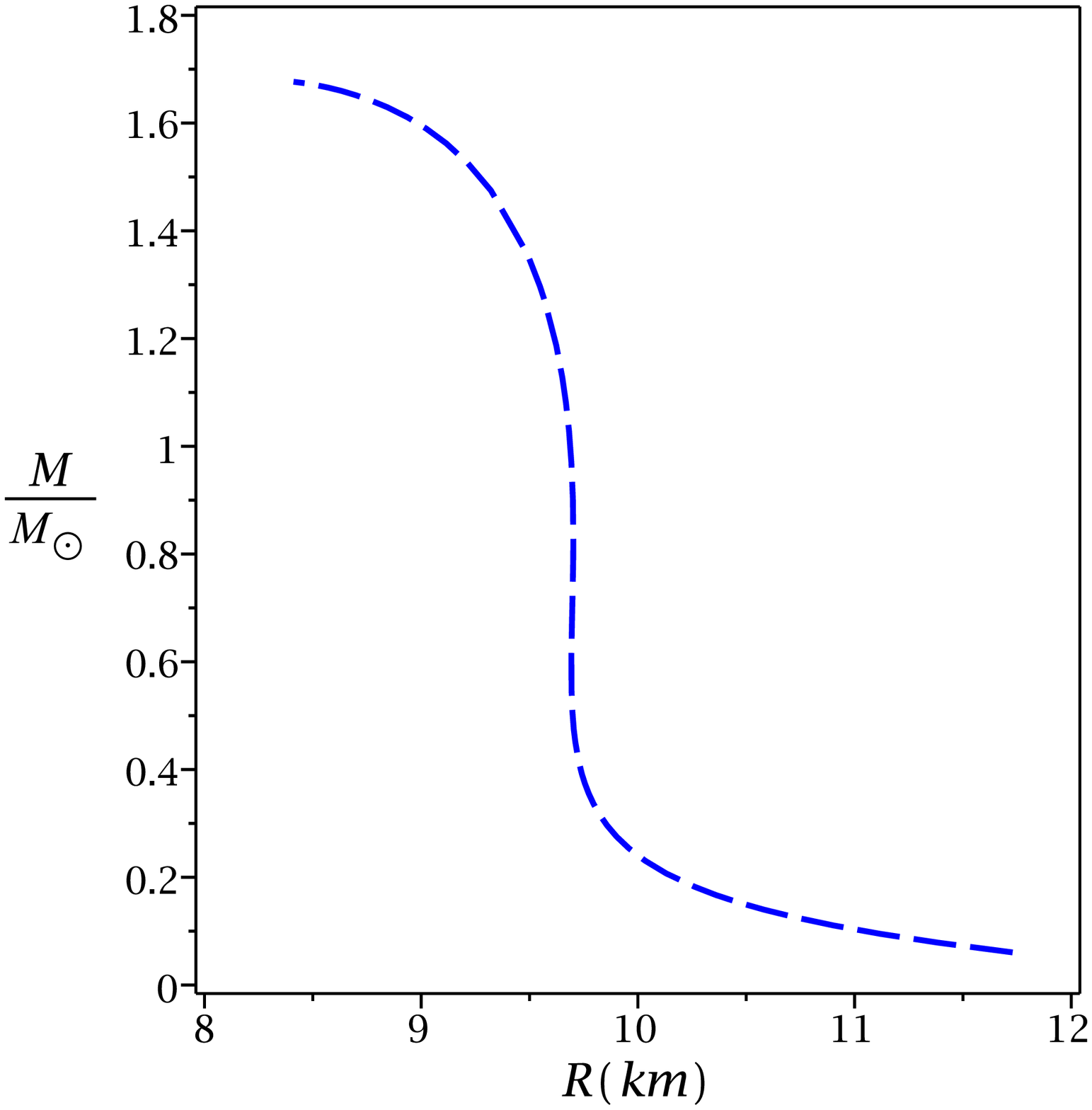} \\
\epsfxsize=7cm \epsffile{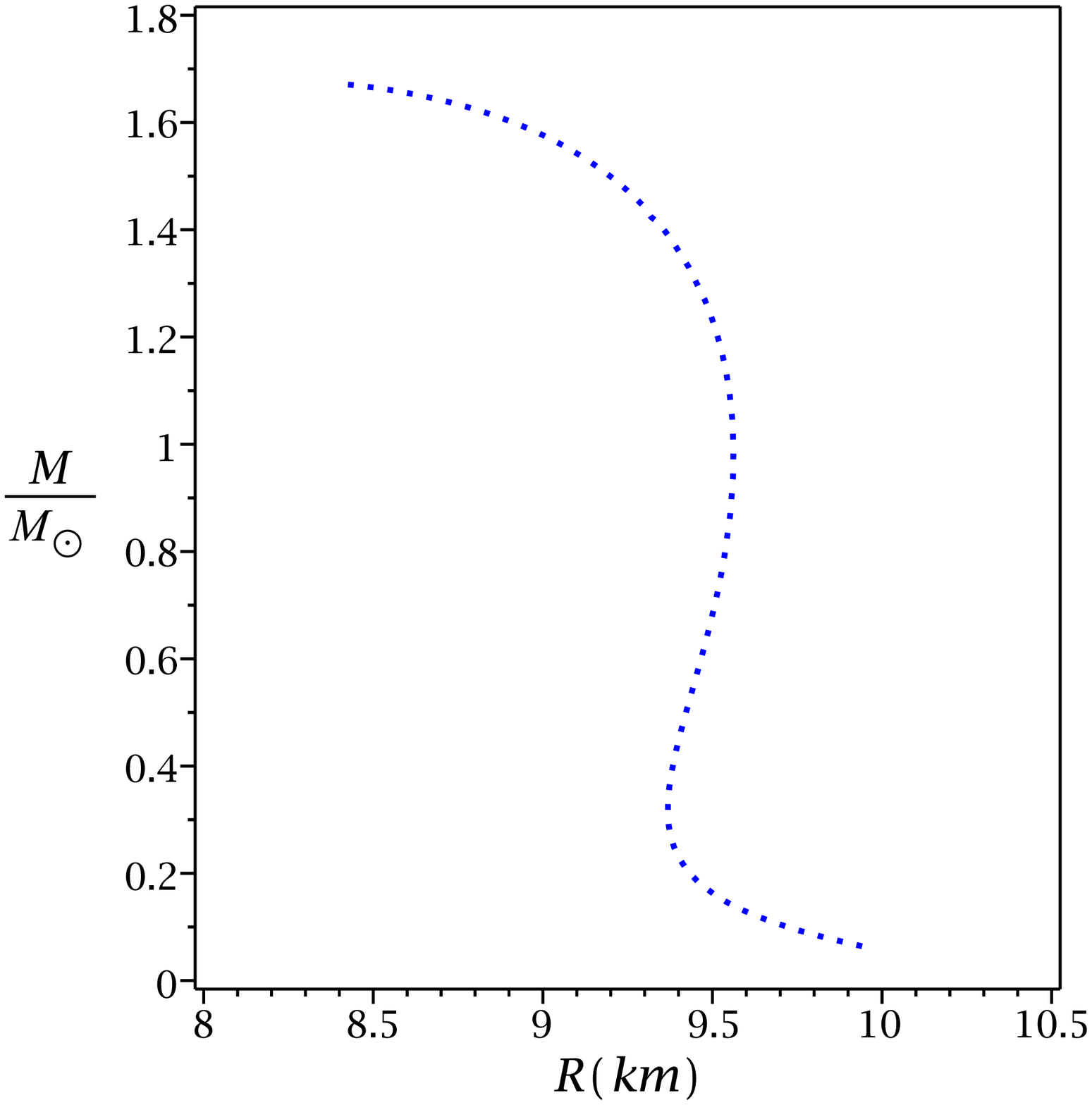} & \epsfxsize=7cm %
\epsffile{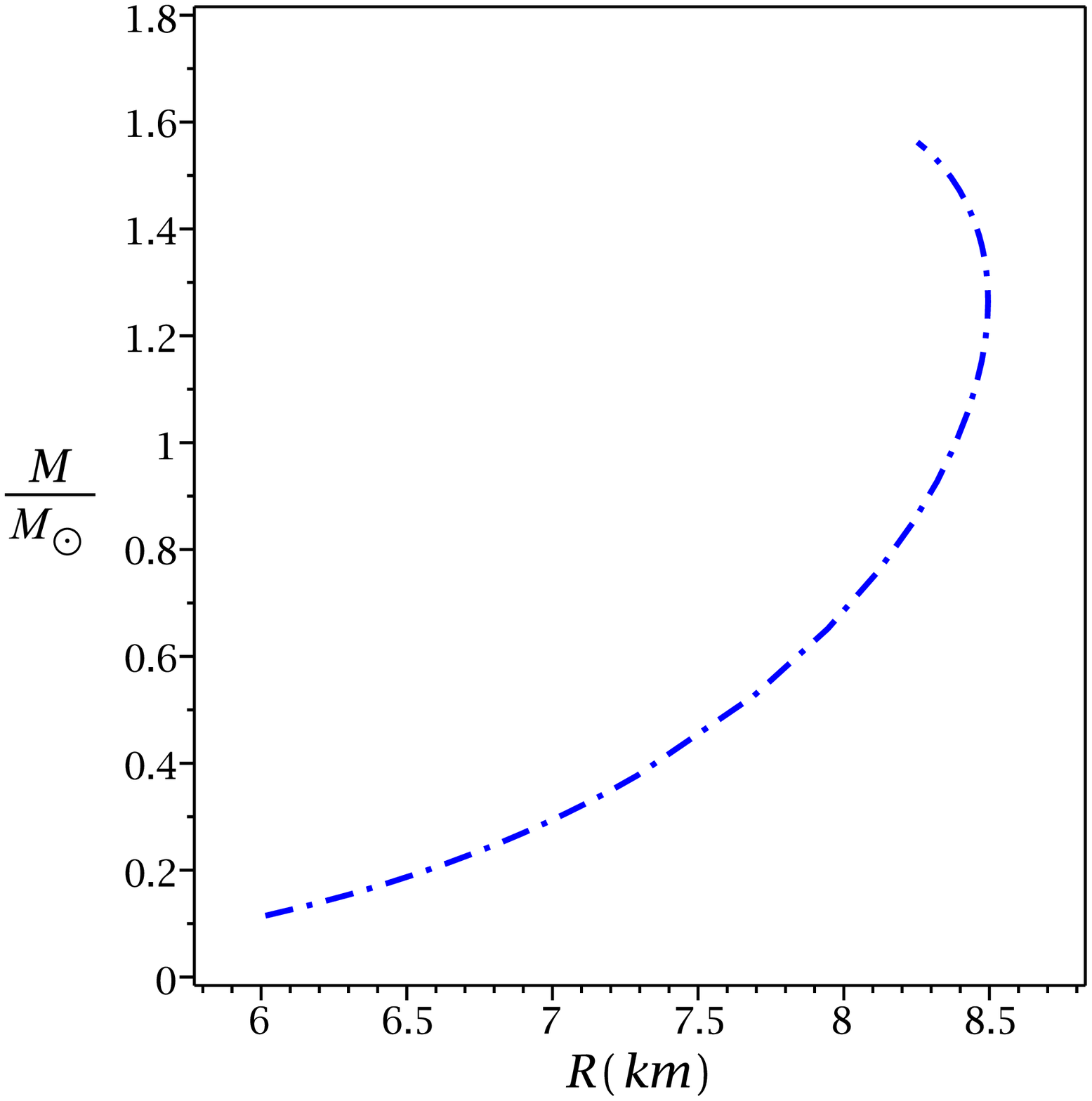}%
\end{array}
$%
\caption{Maximum mass of neutron star versus radius for $\Lambda =1\times
10^{-15}$ (continuous line), $\Lambda =6\times 10^{-14}$ (dashed line), $%
\Lambda =1\times 10^{-13}$ (dotted line) and $\Lambda =1\times 10^{-12}$
(dashed-dotted line).}
\label{Fig3}
\end{figure}

\begin{figure}[tbp]
$%
\begin{array}{c}
\epsfxsize=7cm \epsffile{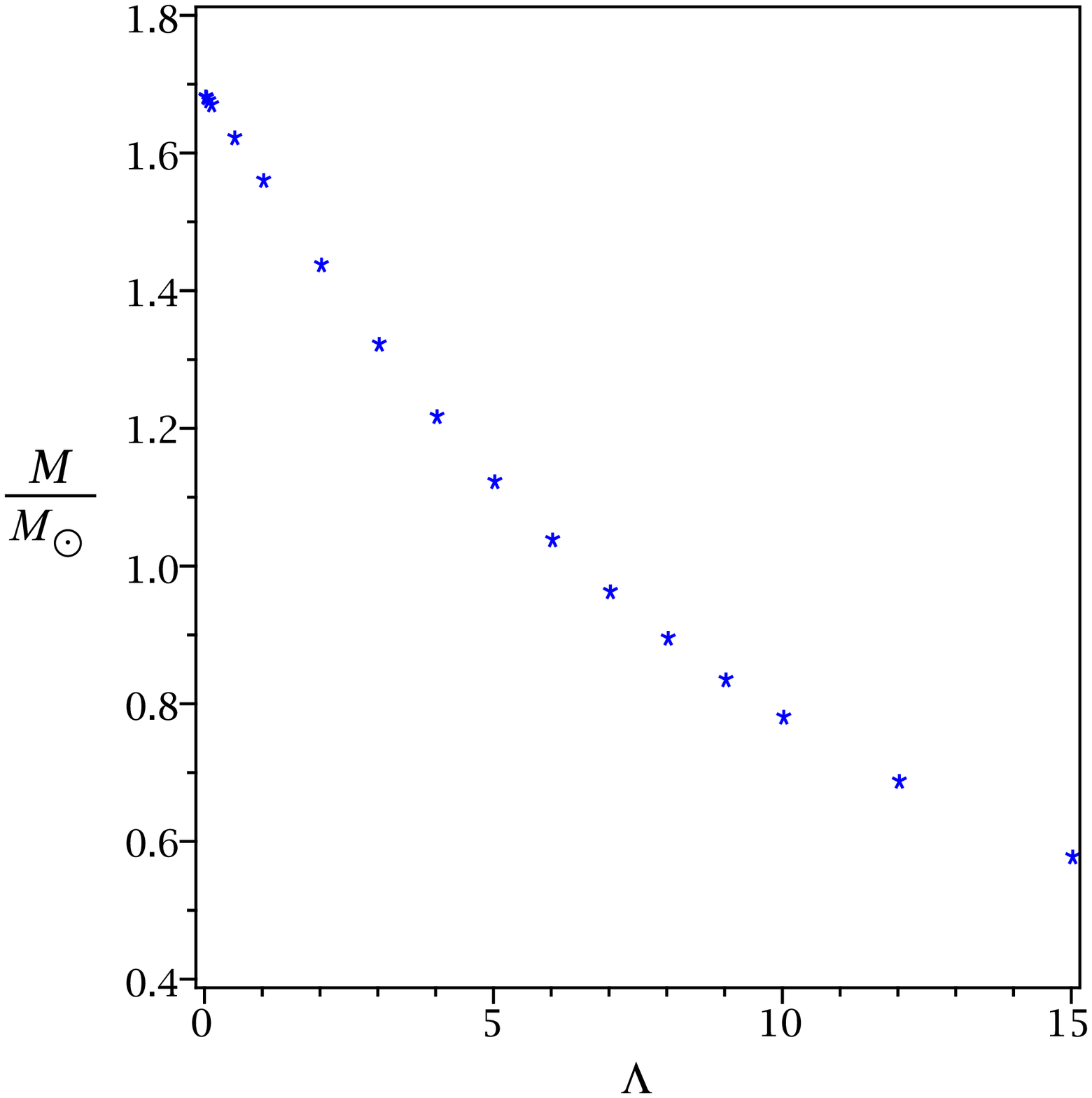}%
\end{array}
$%
\caption{Maximum mass of neutron star versus cosmological constant ($\times
10^{-12}$) in $4$-dimensions.}
\label{Fig4}
\end{figure}

We plot diagrams related to the maximum mass versus central mass density
(radius) in Figs. \ref{Fig2} and \ref{Fig3}. The variation of maximum mass
versus the cosmological constant is also shown in Fig. \ref{Fig4} for $%
\Lambda >0$.

\begin{figure}[tbp]
$%
\begin{array}{c}
\epsfxsize=14cm \epsffile{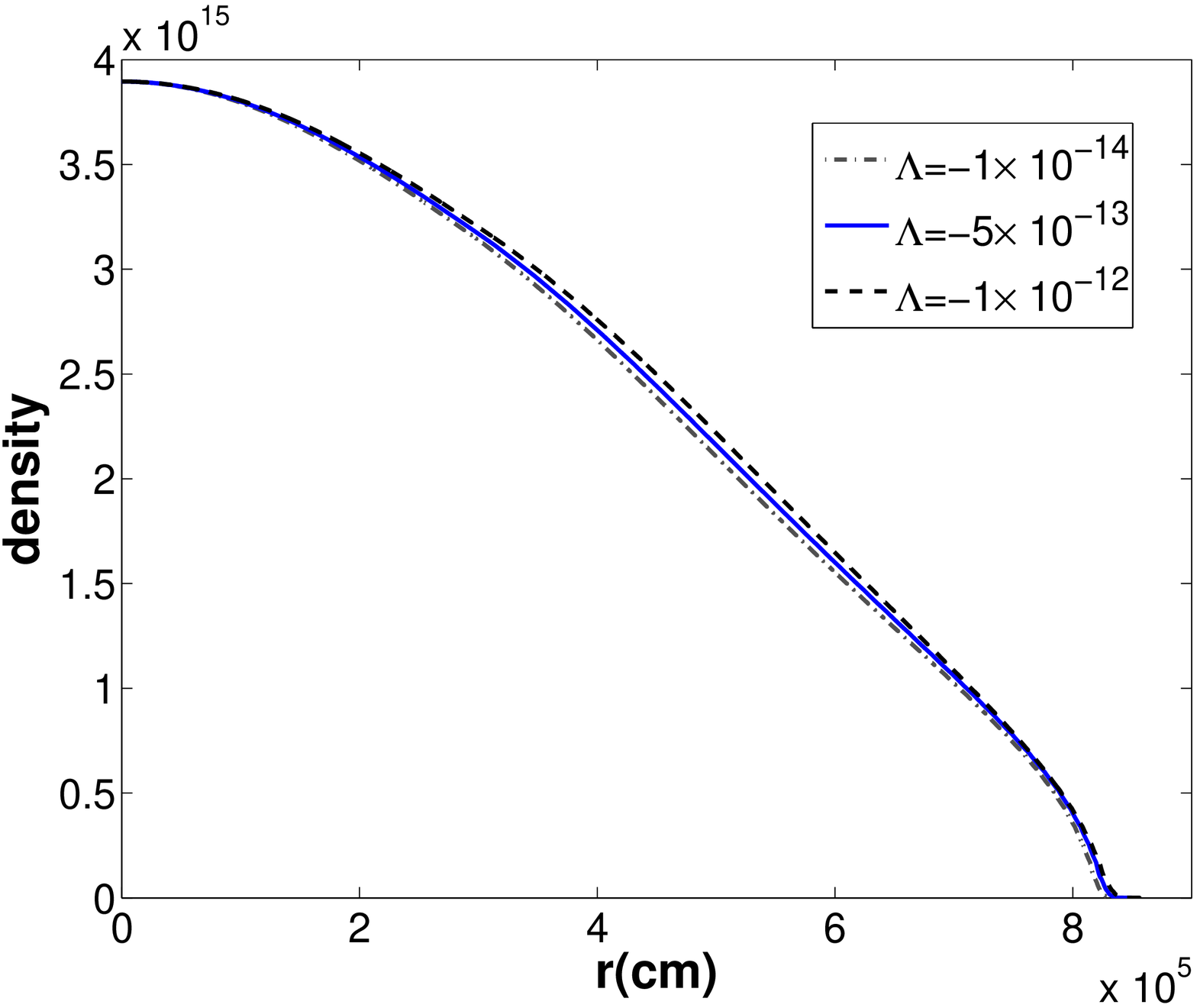}%
\end{array}
$%
\caption{Density of neutron star versus radius.}
\label{Fig44}
\end{figure}

\begin{figure}[tbp]
$%
\begin{array}{c}
\epsfxsize=14cm \epsffile{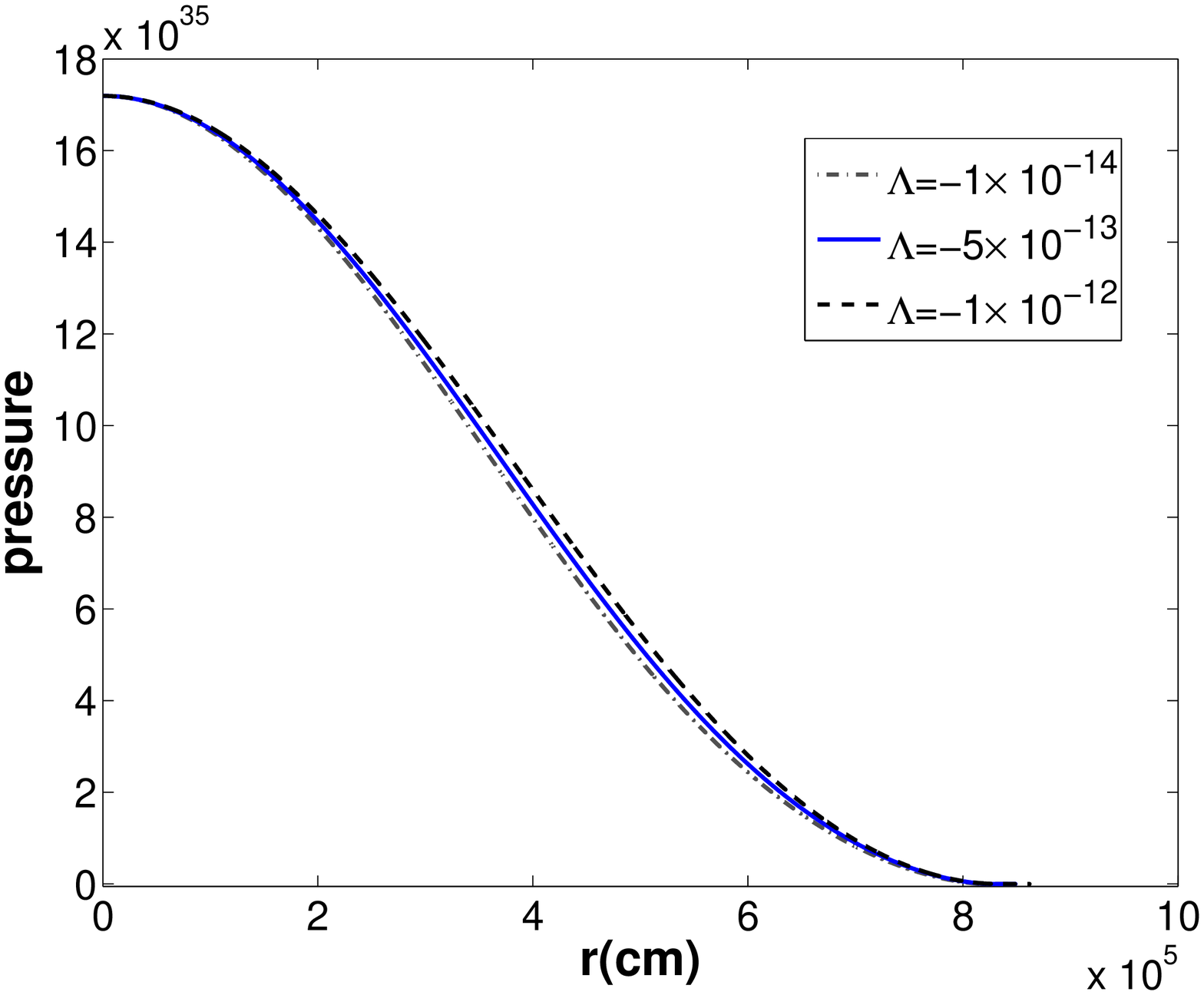}%
\end{array}
$%
\caption{Pressure of neutron star versus radius.}
\label{Fig444}
\end{figure}

\begin{figure}[tbp]
$%
\begin{array}{cc}
\epsfxsize=7cm \epsffile{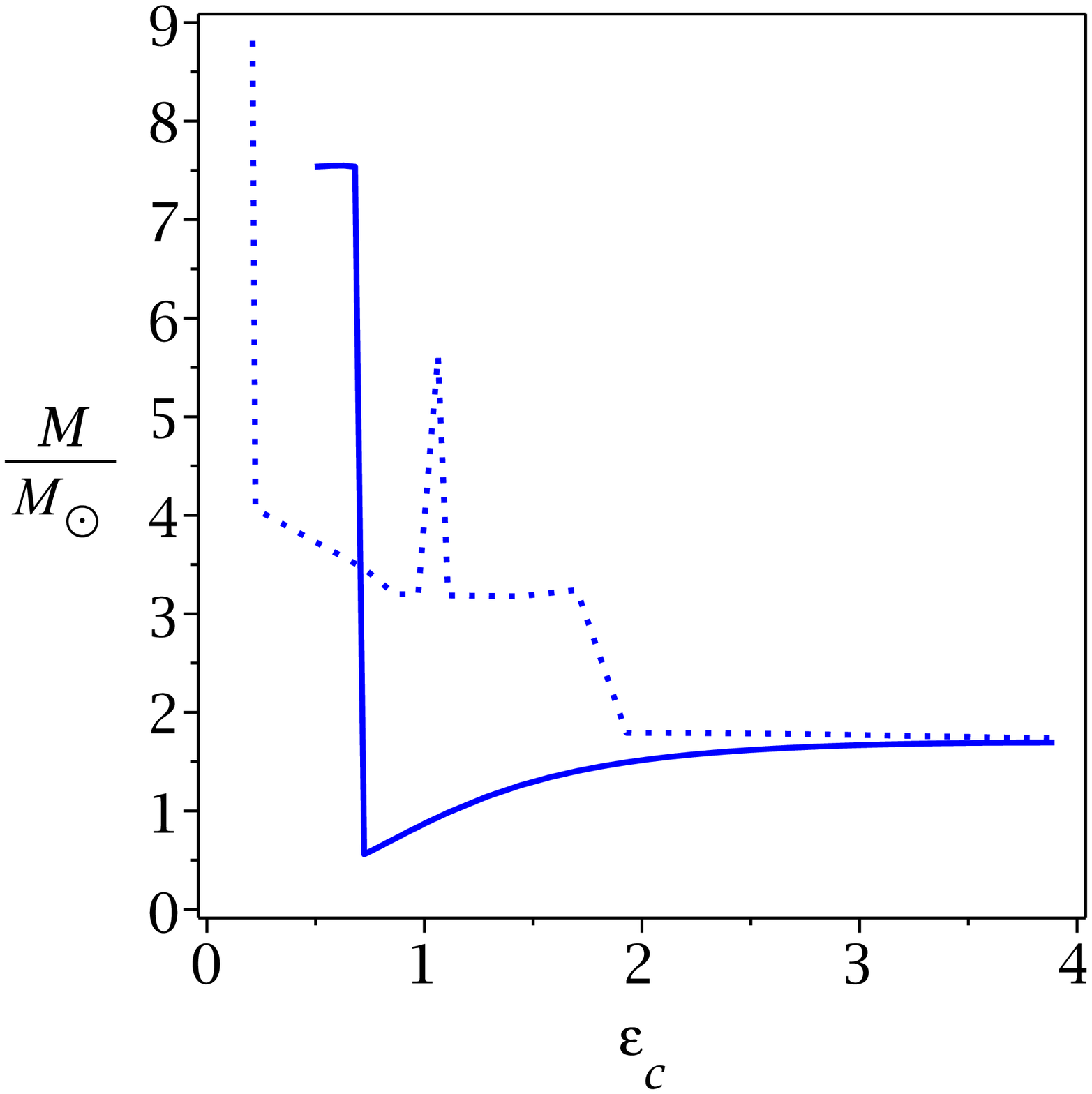} & \epsfxsize=7cm %
\epsffile{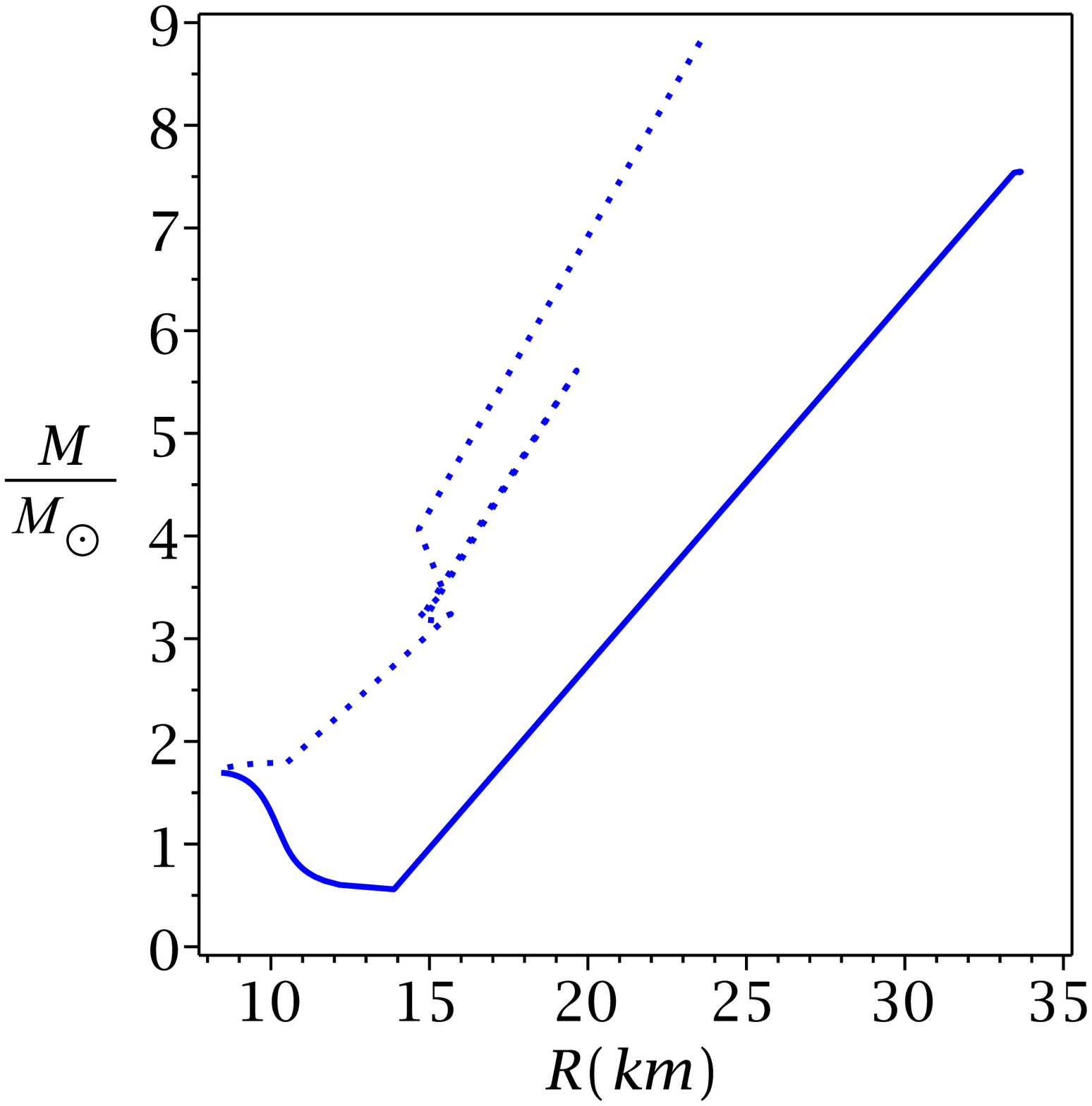}%
\end{array}
$%
\caption{Maximum mass of neutron star versus central mass density $\protect%
\epsilon _{c}$ ($10^{15}$gr/$cm^{3}$), (left panel) and radius (right panel)
for $\Lambda =-9\times 10^{-14}$ (continuous line) and $\Lambda =-5\times
10^{-13}$ (dotted line).}
\label{Fig5}
\end{figure}

As we can see, mass of neutron star versus radius changes when the $\Lambda $
increases. Indeed, one of interesting results of this paper is related to
change in diagrams of mass versus radius, so that by increasing $\Lambda$,
these diagrams change from ordinary shapes of the neutron stars to those of
quark stars (see Fig. \ref{Fig3}). In other words, by increasing $\Lambda$,
we encounter with a change of state (phase transition) of neutron star to
the quasi quark star. On the other hand, we consider negative values of the
cosmological constant and plot the density and pressure versus radius of
neutron star. As one can see, the maximum and minimum values belong to the
center and surface of neutron star, respectively (see Figs. \ref{Fig44} and %
\ref{Fig444}), but the behavior of diagrams related to mass-radius relation
of these stars are not logical (see Fig. \ref{Fig5}). In other words,
Einstein gravity with the negative values of the cosmological constant are
not a suitable theory for explaining the neutron stars with the mentioned
equation of state.

Our results show that, by decreasing the cosmological constant ($\Lambda
<10^{-14}m^{-2}$), the maximum mass and also the radius of this star are not
modified. In other words, the cosmological constant does not significantly
affect the maximum and radius of neutron stars for $\Lambda <10^{-14}\
m^{-2} $. As a result of this paper, when the value of the cosmological
constant is about $10^{-52}~m^{-2}$, this constant does not play a sensible
role in the structure of neutron stars, but by considering larger values for
it (about $\Lambda >10^{-14}\ m^{-2}$), the maximum mass and its radius are
modified (reduced). %

In the following, we investigate other properties of neutron stars by
considering the cosmological constant such as; the Schwarzschild radius,
average density, compactness, Buchdahl-Bondi bound and dynamical stability.

\subsubsection{Schwarzschild Radius}

Now, we obtain the Schwarzschild radius for the obtained masses in Einstein-$%
\Lambda $ gravity in the following form
\begin{equation}
R_{Sch}=\frac{\left[ \left( 3GM+\sqrt{\frac{c^{4}}{\Lambda }+9G^{2}M^{2}}%
\right) \Lambda ^{2}c\right] ^{1/3}}{\Lambda c}-\frac{c}{\left[ \left( 3GM+%
\sqrt{\frac{c^{4}}{\Lambda }+9G^{2}M^{2}}\right) \Lambda ^{2}c\right] ^{1/3}}%
,
\end{equation}%
where $R_{Sch}$ is the Schwarzschild radius, $G=6.67\times 10^{-11}$ $%
Nm^{2}/kg^{2}$ and $c=3\times 10^{8}$ $m/s$. It is notable that, by applying
the cosmological constant to the Einstein gravity, the Schwarzschild radius
modified and also, considering $\Lambda =0$ in Eq. (\ref{Rsch4Lambda}), it
reduces to the Schwarzschild radius in Einstein gravity as \cite%
{Schwarzschild}
\begin{equation}
R_{Sch}=\frac{2GM}{c^{2}}.
\end{equation}

To find the Schwarzschild radius of neutron stars, we use of Eq. (\ref%
{Rsch4Lambda}). The results show that increasing the cosmological constant
and decreasing the maximum mass of neutron stars, the Schwarzschild radius
decreases (see Table \ref{tab3} for more details).

\subsubsection{Average Density}

The average density of the neutron stars in $4-$dimensions can be written as
the following form
\begin{equation}
\overline{\epsilon }=\frac{3M}{4\pi R^{3}},
\end{equation}%
where the results for various the cosmological constant are presented in
Table \ref{tab3}.

\subsubsection{Compactness}

The compactness of a spherical object may be defined as the ratio of the
Schwarzschild radius to the radius of object as
\begin{equation}
\eta =\frac{R_{Sch}}{R},
\end{equation}%
which may be indicated as the strength of gravity. We obtain the values of $%
\eta $ in the Einstein-$\Lambda $ gravity and find that increasing $\Lambda $
leads to decreasing the strength of gravity.

\subsubsection{The Gravitational Redshift}

Using the equation (\ref{4g(r)}), we obtain the gravitational redshift in $4$%
-dimensions for this gravity as
\begin{equation}
z=\frac{1}{\sqrt{1-\frac{2GM}{c^{2}R}-\frac{\Lambda }{3}R^{2}}}-1.
\end{equation}

As one can see, decreasing $\Lambda $ leads to increasing the gravitational
redshift (see Table \ref{tab3}). The maximum value of the gravitational
redshift is about $0.56$. According to Table \ref{tab3}, one finds variation
of $\Lambda$ changes the gravitational redshift, drastically. This is a way
for adjusting the values of $\Lambda$ based on observational evidences.

\subsubsection{Buchdahl-Bondi Bound}

According to Buchdahl-Bondi bound \cite{Buchdahl1,Buchdahl2,Buchdahl3}, for
a static spherically symmetric perfect fluid sphere, maximum permitted $M-R$
ratio should be $M\leq \frac{4c^{2}}{9G}R$. For the Einstein-$\Lambda $
gravity this bound modified as \cite{MakDH}
\begin{equation}
M\leq M_{BB}
\end{equation}%
where $M_{BB}=\frac{2c^{2}}{9G}R-\frac{\Lambda c^{2}}{3G}R^{3}+\frac{2c^{2}}{%
9G}R\sqrt{1+3\Lambda R^{2}}$. Using the above condition, we investigate the
stability condition of neutron stars in Einstein-$\Lambda $ gravity. The
calculations are presented in Table \ref{tab3}. The results show that
obtained maximum mass of neutron stars for all values obtained in Table \ref%
{tab3} are stable.

\subsubsection{Dynamical Stability}

The dynamical stability of the stellar model against the infinitesimal
radial adiabatic perturbation was introduced by Chandrasekhar in Ref. \cite%
{Chandrasekhar}. Next, this stability condition was developed and applied to
astrophysical cases by many authors \cite%
{Bardeen1,Bardeen2,Bardeen3,Bardeen4}. The adiabatic index ($\gamma $) is
defined as
\begin{equation}
\gamma =\frac{\epsilon c^{2}+P}{c^{2}P}\frac{dP}{d\epsilon }.
\label{adiabatic}
\end{equation}

It is notable that, in order to have the dynamical stability, $\gamma $
should be larger than $\frac{4}{3}$ ($\gamma >\frac{4}{3}=1.33$) everywhere
within the isotropic stars. So, we plot a diagram related to $\gamma $
versus radius for different values of the cosmological constant in Fig. \ref%
{adiabaticI}. As one can see, this stellar model is stable against the
radial adiabatic infinitesimal perturbations.

On the other hand, in order to present more investigations for the inside of
neutron stars in Einstein-$\Lambda $ gravity, we plot density and pressure
versus radius of this star in Figs. \ref{density} and \ref{pressure}. These
figures show that density (pressure) values have a maximum and a minimum in
the center and surface of neutron star, respectively, as expected.

\begin{figure}[tbp]
$%
\begin{array}{c}
\epsfxsize=14cm \epsffile{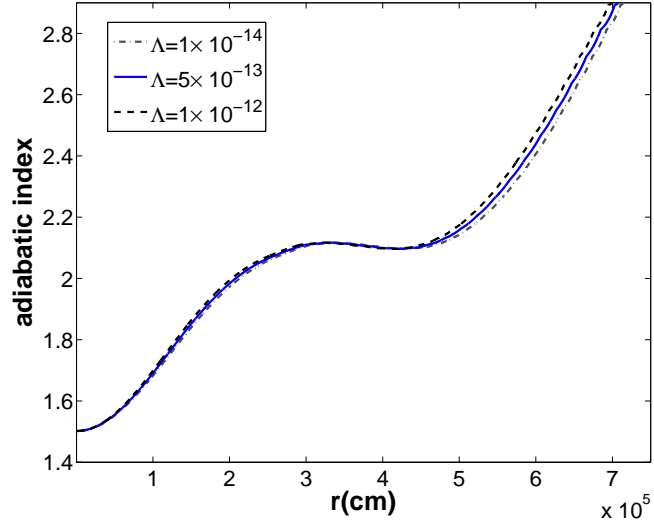}%
\end{array}
$%
\caption{Adiabatic index of neutron star versus radius.}
\label{adiabaticI}
\end{figure}

\begin{figure}[tbp]
$%
\begin{array}{c}
\epsfxsize=14cm \epsffile{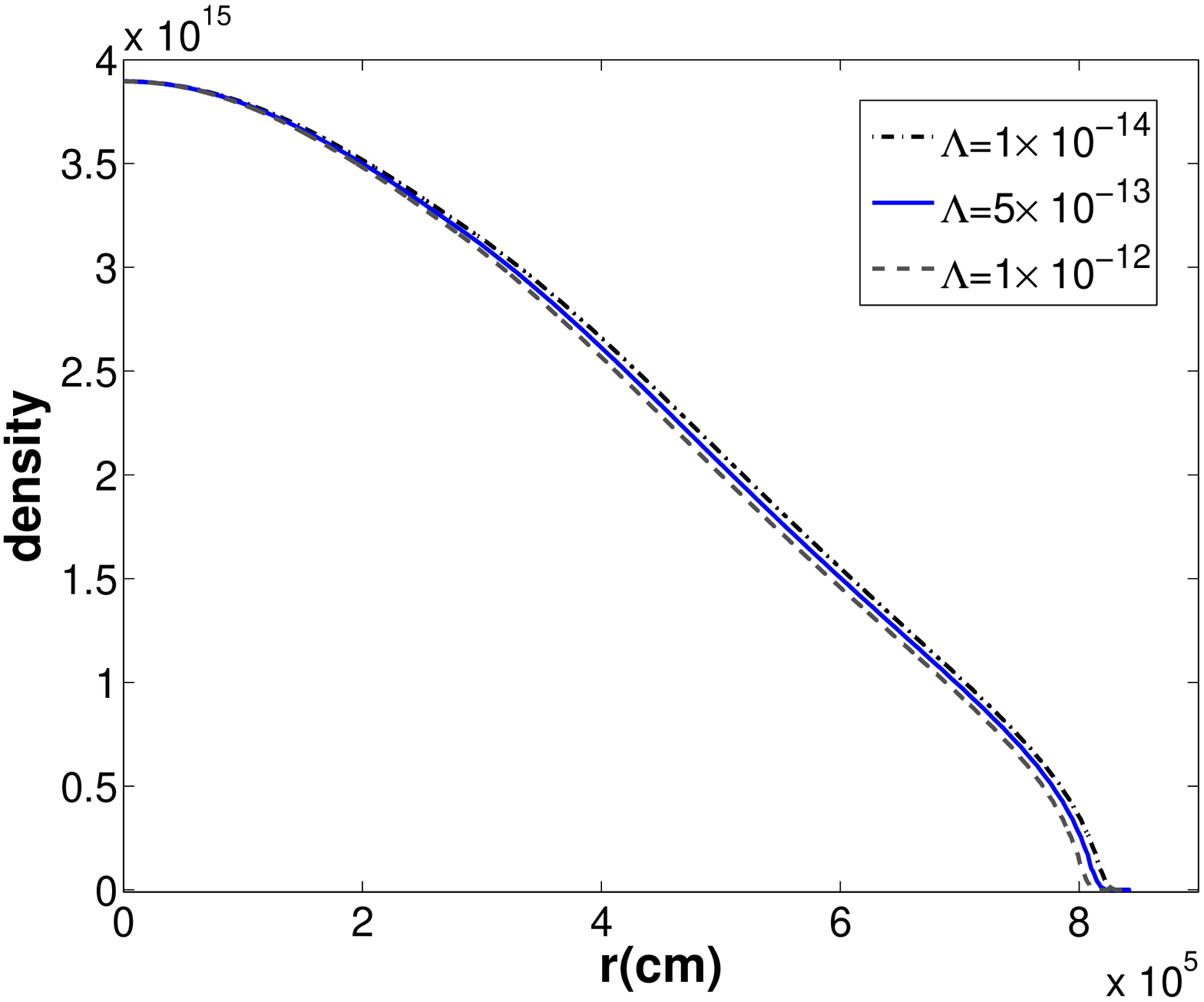}%
\end{array}
$%
\caption{Density of neutron star versus radius.}
\label{density}
\end{figure}

\begin{figure}[tbp]
$%
\begin{array}{c}
\epsfxsize=14cm \epsffile{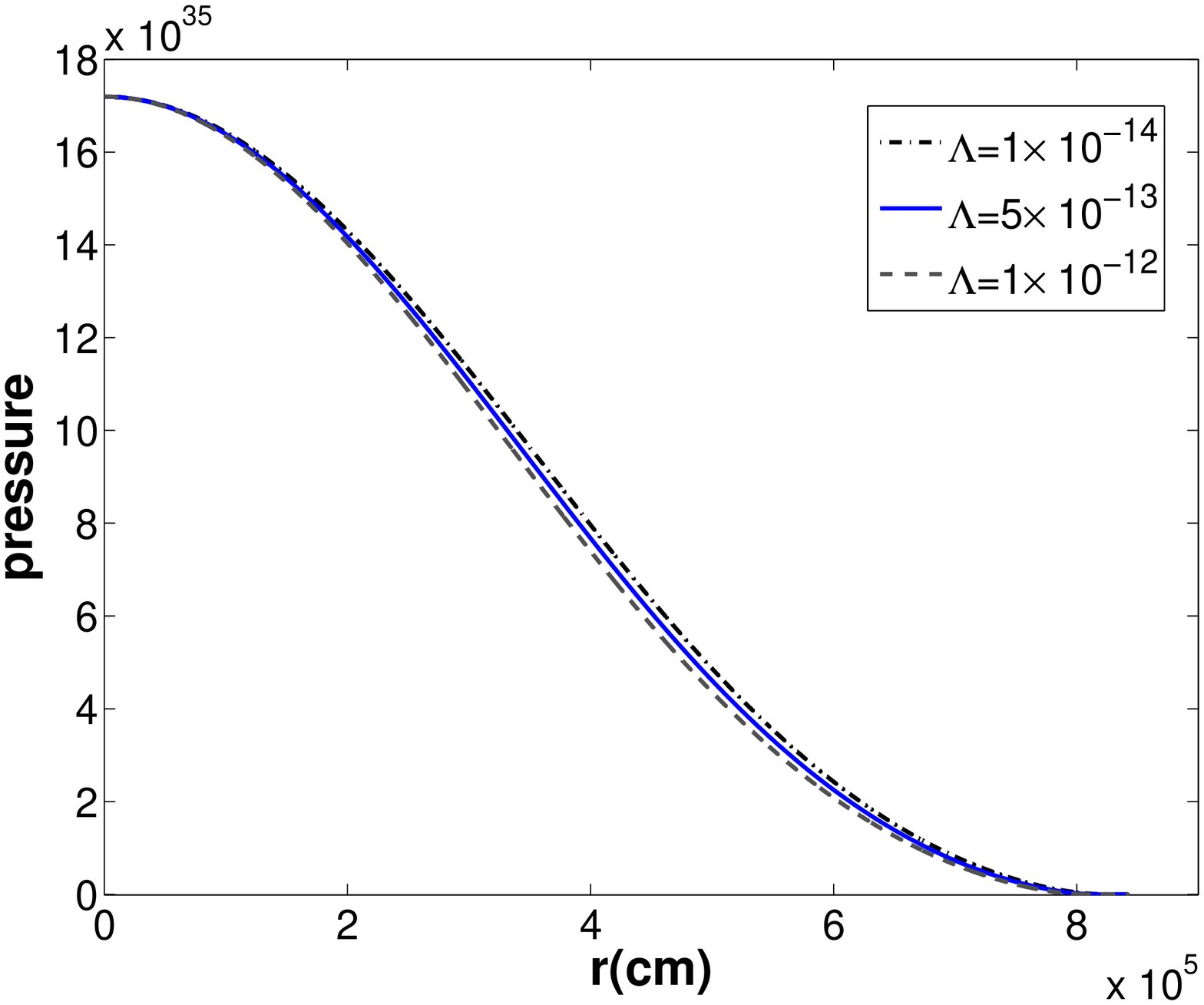}%
\end{array}
$%
\caption{Pressure of neutron star versus radius.}
\label{pressure}
\end{figure}

\section{ Closing Remarks}

In the first part of this paper, we have considered a spherical symmetric
metric and extracted the hydrostatic equilibrium equation of stars for $d-$%
dimensional Einstein-$\Lambda$ gravity ($d\geq 3$). Then, we have
investigated the maximum mass of neutron stars by using the modern equations
of state of neutron star matter derived from microscopic calculations. The
results showed that the cosmological constant affects the maximum mass of
neutron stars. In other words, considering the $\Lambda >0$, the maximum
mass for neutron stars decreases when $\Lambda $ increases. In this case,
maximum mass of neutron stars is in the range $M_{\max }\leq 1.68M_{\odot }$.

One of the interesting results obtained in this paper is the behavior of
diagrams of mass versus radius of the neutron stars. By increasing $\Lambda$%
, these diagrams showed a transition from the neutron star to the quasi
quark star. On the other hand, considering $\Lambda <0$, the behavior of
diagrams related to mass versus radius (central mass density) of neutron
stars is not logical, so the Einstein gravity with the negative values of
cosmological constant can not explain the existence of neutron stars. In
order to investigate the interior of neutron stars in Einstein-$\Lambda $
gravity in more details, we have plotted density and pressure versus radius
of this star. The results showed that the density and pressure were maximum
and minimum of values at the central and surface of neutron stars,
respectively.

Our results showed that, when the approximate value of the cosmological
constant was $10^{-52}~m^{-2}$, this constant did not play a sensitive role
in the structure of neutron stars, but by taking larger values for it (about
$\Lambda >10^{-14}\ m^{-2}$), the maximum mass and its radius were reduced.
Another interesting result of this paper is related to the effect of
cosmological constant on the so-called gravity strength. We have shown that
when the positive values of the cosmological constant increases, the
strength of gravity decreases, and this effect leads to decreasing the
maximum mass of neutron stars.

Finally, we have studied the stability condition for these stars using
Buchdahl-Bondi bound and found that the obtained maximum mass of neutron
stars for all values obtained in Table \ref{tab3} were stable. Then,
considering adiabatic index, we showed that these stars were stable against
the radial adiabatic infinitesimal perturbations.

\begin{center}
\textbf{acknowledgements}
\end{center}

We would like to thank the anonymous referee for his/her valuable comments.
We also wish to thank Shiraz University Research Council. GHB wishes to
thank the Center for Excellence in Astronomy and Astrophysics (CEAA-RIAAM)
for financial support. BE acknowledges S. Panahiyan for helpful discussions.
This work has been supported financially by Research Institute for Astronomy
and Astrophysics of Maragha.


\begin{thebibliography}{999}
\bibitem{Perlmutter1} S. Perlmutter et al., Astrophys. J. \textbf{517}, 565
(1999).

\bibitem{Perlmutter2} S. Perlmutter, M. S. Turner and M. White, Phys. Rev.
Lett. \textbf{83}, 670 (1999).

\bibitem{Perlmutter3} A. G. Riess et al., Astrophys. J. \textbf{607}, 665
(2004).

\bibitem{Lovelock1} D. Lovelock. J. Math. Phys. \textbf{12}, 498 (1971).

\bibitem{Lovelock2} D. Lovelock. J. Math. Phys. \textbf{13}, 874 (1972).

\bibitem{Dehghani1} M. H. Dehghani and S. H. Hendi, Gen. Relativ. Gravit.
\textbf{41}, 1853 (2009).

\bibitem{Dehghani2} S. H. Hendi, B. Eslam Panah and S. Panahiyan, Phys. Rev.
D \textbf{91}, 084031 (2015).

\bibitem{Brax1} P. Brax and C. van de Bruck, Class. Quantum Gravit. \textbf{%
20}, R201 (2003).

\bibitem{Brax2} L. A. Gergely, Phys. Rev. D \textbf{74}, 024002 (2006).

\bibitem{Brax3} M. Demetrian, Gen. Relativ. Gravit. \textbf{38}, 953 (2006).

\bibitem{Brans1} C. Brans and R. H. Dicke, Phys. Rev. \textbf{124}, 925
(1961).

\bibitem{Brans2} R. G. Cai and Y. S. Myung, Phys. Rev. D \textbf{56}, 3466
(1997).

\bibitem{Brans3} T. P. Sotiriou, Class. Quantum Gravit. \textbf{23}, 5117
(2006).

\bibitem{Brans4} K. I. Maeda and Y. Fujii, Phys. Rev. D \textbf{79}, 084026
(2009).

\bibitem{Brans5} S. H. Hendi and R. Katebi, Eur. Phys. J. C \textbf{72}, 1
(2012).

\bibitem{Harada1} T. Harada, Phys. Rev. D \textbf{57}, 4802 (1998).

\bibitem{Harada2} D. D. Doneva, S. S. Yazadjiev, N. Stergioulas, K. D.
Kokkotas and T. M. Athanasiadis, Phys. Rev. D \textbf{90}, 044004 (2014).

\bibitem{Harada3} M. Ponce, C. Palenzuela, E. Barausse and L. Lehner, Phys.
Rev. D \textbf{91}, 084038 (2015).

\bibitem{Harada4} A. Cisterna, T. Delsate and M. Rinaldi, Phys. Rev. D
\textbf{92}, 044050 (2015).

\bibitem{Harada5} C. Palenzuela and S. Liebling, Phys. Rev. D \textbf{93},
044009 (2016).

\bibitem{Souza1} J. C. C. de Souza and V. Faraoni. Class. Quantum Gravit.
\textbf{24}, 3637 (2007).

\bibitem{Souza2} G. Cognola, E. Elizalde, S. Nojiri, S. D. Odintsov, L.
Sebastiani, and S. Zerbini. Phys. Rev. D, \textbf{77}, 046009 (2008)

\bibitem{Souza3} S. Nojiri and S. D. Odintsov, Phys. Rept. \textbf{505}, 59
(2011).

\bibitem{Souza4} N. R. Napolitano, S. Capozziello, A. J. Romanowsky, M.
Capaccioli and C. Tortora, Astrophys. J. \textbf{748}, 87 (2012).

\bibitem{Souza5} S. H. Hendi, R. B. Mann, N. Riazi and B. Eslam Panah, Phys.
Rev. D \textbf{86}, 104034 (2012).

\bibitem{Souza6} M. Lubini, C. Tortora, J. Naf, Ph. Jetzer and S.
Capozziello, Eur. Phys. J. C \textbf{71}, 1834 (2011).

\bibitem{Cooney1} A. Cooney, S. D. Deo and D. Psaltis, Phys. Rev. D \textbf{%
82}, 064033 (2010).

\bibitem{Cooney2} S. H. Hendi, B. Esam Panah and S. M. Mousavi, Gen.
Relativ. Gravit. \textbf{44}, 835 (2012).

\bibitem{Cooney3} S. H. Hendi, B. Eslam Panah and C. Corda, Can. J. Phys.
\textbf{92}, 76 (2014).

\bibitem{Cooney4} S. Capozziello, M. D. Laurentis, R. Farinelli and S. D.
Odintsov, Phys. Rev. D \textbf{93}, 023501 (2016).

\bibitem{Cooney5} C. Gao and Y. G. Shen, Gen. Relativ. Gravit. \textbf{48},
1 (2016).

\bibitem{Peebles} P. J. E. Peebles and B. Ratra, Rev. Mod. Phys. \textbf{75}%
, 559 (2003).

\bibitem{Padmanabhan1} T. Padmanabhan, Phys. Rept. \textbf{380}, 235 (2003).

\bibitem{Padmanabhan2} J. A. Frieman, M. S. Turner and D. Huterer, Ann. Rev.
Astron. Astrophys. \textbf{46}, 385 (2008).

\bibitem{TOV1} R. C. Tolman, Proc. Nat. Acad. Sc. \textbf{20}, 169 (1934).

\bibitem{TOV2} R. C. Tolman, Phys. Rev. \textbf{55}, 364 (1939).

\bibitem{TOV3} J. R. Oppenheimer and G. M. Volkoff, Phys. Rev. \textbf{55},
374 (1939).

\bibitem{Silbar1} N. Yunes and M. Visser, Int. J. Mod. Phys. A \textbf{18},
3433 (2003).

\bibitem{Silbar2} R. R. Silbar and S. Reddy, Am. J. Phys. \textbf{72}, 892
(2004).

\bibitem{Silbar3} G. Narain, J. Schaffner-Bielich and I. N. Mishustin, Phys.
Rev. D \textbf{74}, 063003 (2006).

\bibitem{Silbar4} G. H. Bordbar, M. Bigdeli and T. Yazdizade, Int. J. Mod.
Phys. A \textbf{21}, 5991 (2006).

\bibitem{Silbar5} P. Boonserm, M. Visser and S. Weinfurtner, Phys. Rev. D
\textbf{76}, 044024 (2007).

\bibitem{Silbar6} X. Li, F. Wang and K. S. Cheng, JCAP \textbf{10}, 031
(2012).

\bibitem{Silbar7} A. M. Oliveira, H. E. S. Velten, J. C. Fabris and I. G.
Salako, Eur. Phys. J. C \textbf{74}, 3170 (2014).

\bibitem{Silbar8} X. T. He, F. J. Fattoyev, B. A. Li and W. G. Newton, Phys.
Rev. C \textbf{91}, 015810 (2015).

\bibitem{Various gravity1} T. Wiseman, Phys. Rev. D \textbf{65}, 124007
(2002).

\bibitem{Various gravity2} H. Sotani, Phys. Rev. D \textbf{86}, 124036
(2012).

\bibitem{Various gravity3} D. D. Doneva, S. S. Yazadjiev, N. Stergioulas and
K. D. Kokkotas, Phys. Rev. D \textbf{88}, 084060 (2013).

\bibitem{Various gravity4} N. Chamel, P. Haensel, J. L. Zdunik and A. F.
Fantina, Int. J. Mod. Phys. E \textbf{22}, 1330018 (2013).

\bibitem{Various gravity5} E. Barausse, C. Palenzuela, M. Ponce and L.
Lehner, Phys. Rev. D \textbf{87}, 081506(R) (2013).

\bibitem{Various gravity6} H. O. Silva, H. Sotani, E. Berti and M.
Horbatsch, Phys. Rev. D \textbf{90}, 124044 (2014).

\bibitem{Various gravity7} Y. Brihaye and J. Riedel, Phys. Rev. D \textbf{89}%
, 104060 (2014).

\bibitem{Various gravity8} H. O. Silva, C. F. B. Macedo, E. Berti and L. C.
B. Crispino, Class. Quantum Gravit. \textbf{32}, 145008 (2015).

\bibitem{Various gravity9} A. Das, F. Rahaman, B. K. Guha and S. Ray,
Astrophys. Space. Sci. \textbf{358}, 36 (2015).

\bibitem{Various gravity10} K. V. Staykov, D. D. Doneva, S. S. Yazadjiev and
K. D. Kokkotas, Phys. Rev. D \textbf{92}, 043009 (2015).

\bibitem{Hendi2015} S. H. Hendi, G. H. Bordbar, B. Eslam Panah and M.
Najafi, Astrophys. Space. Sci. \textbf{358}, 30 (2015).

\bibitem{HendiBEP} S. H. Hendi, G. H. Bordbar, B. Eslam Panah and S.
Panahiyan, JCAP \textbf{09}, 013 (2016).

\bibitem{HendiBEmassive} T. Katsuragawa, S. Nojiri, S. D. Odintsov, M.
Yamazaki, Phys. Rev. D \textbf{93}, 124013 (2016).

\bibitem{Astashenok2015a} A. V. Astashenok, S. Capozziello and S. D.
Odintsov, Astrophys. Space. Sci. \textbf{355}, 341 (2015).

\bibitem{Astashenok2015b} A. V. Astashenok, S. Capozziello and S. D.
Odintsov, JCAP \textbf{01}, 001 (2015).

\bibitem{Momeni1} G. Abbas, D. Momeni, M. Amir Ali, R. Myrzakulov and S.
Qaisar, Astrophys. Space Sci. \textbf{357}, 158 (2015).

\bibitem{Momeni2} S. S. Yazadjiev, D. D. Doneva and K. D. Kokkotas, Phys.
Rev. D \textbf{91}, 084018 (2015).

\bibitem{Momeni3} A. Savas Arapoglu, C. Deliduman and K. Yavuz Eksi, JCAP
\textbf{07}, 020 (2011).

\bibitem{Momeni4} K. Zhou, Z. Y. Yang, D. C. Zou and R. H. Yue, Chin. Phys.
B \textbf{21}, 020401 (2012).

\bibitem{Glampedakis} K. Glampedakis, G. Pappas, H. O. Silva and E. Berti,
Phys. Rev. D \textbf{92}, 024056 (2015).

\bibitem{Boyadjiev} T. Boyadjiev, P. Fiziev and S. Yazadjiev, Class. Quantum
Gravit. \textbf{16}, 2359 (1999).

\bibitem{Vinayaraj1} O. K. Vinayaraj and V. C. Kuriakose, [arXiv:0802.1155].

\bibitem{Vinayaraj2} M. Heydari-Fard and H. R. Sepangi, JCAP \textbf{02},
029 (2009).

\bibitem{Vinayaraj3} S. Meyer, F. Pace and M. Bartelmann, Phys. Rev. D
\textbf{86}, 103002 (2012).

\bibitem{Astashenok2013a} A. V. Astashenok, S. Capozziello and S. D.
Odintsov, JCAP \textbf{12}, 040 (2013).

\bibitem{Astashenok2013b} M. Orellana, F. Garc\'{\i}a, F. A. T. Pannia and
G. E. Romero, Gen. Relativ. Gravit. \textbf{45}, 771 (2013).

\bibitem{Astashenok2013c} J. D. V. Arbanil, J. P. S. Lemos and V. T.
Zanchin, Phys. Rev. D \textbf{88}, 084023 (2013).

\bibitem{Astashenok2013d} R. Goswami, A. M. Nzioki, S. D. Maharaj and S. G.
Ghosh, Phys. Rev. D \textbf{90}, 084011 (2014).

\bibitem{Lemos1} J. P. S. Lemos, F. J. Lopes, G. Quinta and V. T. Zanchin,
Eur. Phys. J. C \textbf{75}, 76 (2015).

\bibitem{Lemos2} D. Momeni, H. Gholizade, M. Raza and R. Myrzakulov, Int. J.
Mod. Phys. A \textbf{30}, 1550093 (2015).

\bibitem{Lemos3} K. Glampedakis, G. Pappas, H. O. Silva and E. Berti, Phys.
Rev. D \textbf{92}, 024056 (2015).

\bibitem{Lemos4} H. Velten, A. M. Oliveira and A. Wojnar, [arXiv:1601.03000].

\bibitem{Green1} M. B. Green and J. H. Schwarz, Phys. Lett. B \textbf{151},
21 (1985).

\bibitem{Green2} P. Candelas, G. Horowitz, A. Strominger and E. Witten,
Nucl. Phys. B \textbf{258}, 46 (1985).

\bibitem{Diaz} A. A. G. Diaz, [arXiv:1412.5620].

\bibitem{Shapiro} S. Shapiro and S. Teukolsky, \textit{Black Holes, White
Dwarfs and Neutron Stars}. Wiley, New York (1983).

\bibitem{Bordbar1} G. H. Bordbar and N. Riazi, Astrophys. Space Sci. \textbf{%
282}, 563 (2002).

\bibitem{Bordbar2} G. H. Bordbar, Int. J. Theor. Phys. \textbf{43}, 399
(2004).

\bibitem{Wiringa} R. B. Wiringa, V. Stoks and R. Schiavilla, Phys. Rev. C
\textbf{51}, 38 (1995).

\bibitem{Stoks} V. G. J. Stoks, R. A. M. Klomp, C. P. F. Terheggen and J. J.
de Swart, Phys. Rev. C \textbf{49}, 2950 (1994).

\bibitem{Bordbar1998} G. H. Bordbar and M. Modarres, Phys. Rev. C \textbf{57}%
, 714 (1998).

\bibitem{Modarres} M. Modarres and G. H. Bordbar, Phys. Rev. C \textbf{58},
2781 (1998).

\bibitem{ModarresI} M. Modarres and J. M. Irvine, J. Phys. G: Nucl. Part.
Phys. \textbf{5}, 511 (1979).

\bibitem{Howes} C. Howes, R. F. Bishop and J. M. Irvine, J. Phys. G: Nucl.
Part. Phys. \textbf{4}, 89 (1978).

\bibitem{ModarresII} M. Modarres, J. Phys. G: Nucl. Part. Phys. \textbf{21},
351 (1995).

\bibitem{Owen} J. C. Owen, R. F. Bishop and J. M. Irvine, Nucl. Phys. A
\textbf{277}, 45 (1977).

\bibitem{BordbarM1997} G. H. Bordbar and M. Modarres, J. Phys. G: Nucl.
Part. Phys. \textbf{23}, 1631 (1997).

\bibitem{Jastrow} J. W. Clark, Prog. Part. Nucl. Phys. \textbf{2}, 89 (1979).

\bibitem{Bordbar044310} G. H. Bordbar, Z. Rezaei and A. Montakhab, Phys.
Rev. C \textbf{83}, 044310 (2011).

\bibitem{BordbarB1} G. H. Bordbar and M. Bigdeli, Phys. Rev. C \textbf{75},
045804 (2007).

\bibitem{BordbarB2} G. H. Bordbar and M. Bigdeli, Phys. Rev. C \textbf{77},
015805 (2008).

\bibitem{Herrera1} L. Herrera, Phys. Lett. A \textbf{165}, 206 (1992).

\bibitem{Herrera2} H. Abreu, H. Hernandez and L. A. Nunes, Class. Quantum
Gravit. \textbf{24}, 4631 (2007).

\bibitem{Bordbar2006} G. H. Bordbar and M. Hayati, Int. J. Mod. Phys. A
\textbf{21}, 1555 (2006).

\bibitem{WiringaFF} R. B. Wiringa, V. Fiks and A. Fabrocini, Phys. Rev. C
\textbf{38}, 1010 (1988).

\bibitem{Weisberg1} J. M. Weisberg and J. H. Taylor, Phys. Rev. Lett.
\textbf{52}, 1348 (1984).

\bibitem{Weisberg2} E. P. Liang, Astrophys. J. \textbf{304}, 682 (1986).

\bibitem{Weisberg3} S. R. Heap and M. F. Corcoran, Astrophys. J. \textbf{387}%
, 340 (1992).

\bibitem{Weisberg4} H. Quaintrell, Astron. Astrophys. \textbf{401}, 313
(2003).

\bibitem{Schwarzschild} K. Schwarzschild, \textit{translation by:} S. Antoci
and A. Loinger, [arXiv:physics/9905030].

\bibitem{Buchdahl1} H. A. Buchdahl, Phys. Rev. \textbf{116}, 1027 (1959).

\bibitem{Buchdahl2} H. Bondi, Proc. R. Soc. Lond. A \textbf{282}, 303 (1964).

\bibitem{Buchdahl3} H. A. Buchdahl, Astrophys. J. \textbf{146}, 275 (1966).

\bibitem{MakDH} M. K. Mak, Peter N. Dobson, Jr., T. Harko, Mod. Phys. Lett.
A \textbf{15}, 2153 (2000).

\bibitem{Chandrasekhar} S. Chandrasekhar, Astrophys. J. \textbf{140}, 417
(1964).

\bibitem{Bardeen1} J. M. Bardeen, K. S. Thonre and D. W. Meltzer, Astrophys.
J. \textbf{145}, 505 (1966).

\bibitem{Bardeen2} H. Kuntsem, MNRAS. \textbf{232}, 163 (1988).

\bibitem{Bardeen3} M. K. Mak and T. Harko, Eur. Phys. J. C \textbf{73}, 2585
(2013).

\bibitem{Bardeen4} M. Kalam, S. M. Hossein and S. Molla, [arXiv:1510.07015].
\end{thebibliography}
\end{document}